\documentclass[conference]{IEEEtran}
\IEEEoverridecommandlockouts
\usepackage{cite}
\usepackage{amsmath,amssymb,amsfonts}
\usepackage{graphicx}
\usepackage{xcolor}
\usepackage{subcaption}
\usepackage{booktabs}

\newcommand{\ie}{i.e.\ }
\usepackage{comment}
\usepackage{svg}

\DeclareMathOperator*{\argmin}{arg\,min}

\def\BibTeX{{\rm B\kern-.05em{\sc i\kern-.025em b}\kern-.08em
    T\kern-.1667em\lower.7ex\hbox{E}\kern-.125emX}}
\begin{document}

\title{Global Speed-of-Sound Prediction\\ Using Transmission Geometry
\thanks{Funding provided by Uppsala Medtech Science and Innovation Centre.}
}

\author{Can Deniz Bezek\textsuperscript{1}, Mert Bilgin\textsuperscript{2}, Lin Zhang\textsuperscript{2}, Orcun Goksel\textsuperscript{1, 2}
\\
\textsuperscript{1} Department of Information Technology, Uppsala University, Uppsala, Sweden
\\
\textsuperscript{2} Department of Information Technology and Electrical Engineering, ETH Zurich, Zurich, Switzerland}

\maketitle

\begin{abstract}
Most ultrasound (US) imaging techniques use spatially-constant speed-of-sound (SoS) values for beamforming. 
Having a discrepancy between the actual and used SoS value leads to aberration artifacts, e.g., reducing the image resolution, which may affect diagnostic usability. 
Accuracy and quality of different US imaging modalities, such as tomographic reconstruction of local SoS maps, also depend on a good initial beamforming SoS. 
In this work, we develop an analytical method for estimating mean SoS in an imaged medium.
We show that the relative shifts between beamformed frames depend on the SoS offset and the geometric disparities in transmission paths. 
Using this relation, we estimate a correction factor and hence a corrected mean SoS in the medium. 
We evaluated our proposed method on a set of numerical simulations, demonstrating its utility both for global SoS prediction and for local SoS tomographic reconstruction. 
For our evaluation dataset, for an initial SoS under- and over-assumption of 5\% the medium SoS, our method is able to predict the actual mean SoS within 0.3\% accuracy.
For the tomographic reconstruction of local SoS maps, the reconstruction accuracy is improved on average by 78.5\% and 87\%, respectively, compared to an initial SoS under- and over-assumption of 5\%. 
\end{abstract}

\begin{IEEEkeywords}
Beamforming, aberration correction.
\end{IEEEkeywords}

\section{Introduction}
Speed-of-Sound (SoS) is the longitudinal travel rate of the acoustic waves within the tissue. 
\emph{Beamforming} is the reconstruction of spatial US images from temporal signals received by an US transducer, and this process requires an SoS assumption for such image formation. 
Typically a constant SoS value is assumed and used. 
However, SoS may in reality change largely in tissues.
Having a discrepancy between the actual and assumed SoS values leads to \textit{aberration} artifacts, e.g., reducing the image resolution, which may affect diagnostic usability. 
Global average SoS values are also utilized in the beamforming stage of tomographic reconstruction of local SoS maps~\cite{rau_2021}.
Thus, utilizing a correct SoS assumption is of great importance in US image formation.
SoS estimation has been studied by several groups, to find either a single global SoS value or a spatial (heterogeneous) SoS map. 

Among single-value estimation methods, in~\cite{anderson_1998} a least-squares fitting of a second-order polynomial to echo profiles was proposed for estimating a mean SoS within the imaged region. 
This work was later extended for measuring SoS in layered phantoms in~\cite{byram_2012}. 
In~\cite{shen_2020}, mean SoS was estimated using the signal coherence between different transmit (Tx) and receive (Rx) paths in multi-angle plane wave imaging. 
In \cite{xenia_2021}, an approach for mean SoS estimation was proposed by exploiting geometric disparities between sequences originating from different transducer elements.
This used calibrations from simulations, in order to map the lateral disparity profiles to actual SoS under-/over-estimation.
Herein we propose an analytical estimation method with a closed form solution, where under-/over-estimation can be directly inferred from disparities, \ie without a need for calibration, in contrast to~\cite{xenia_2021}.

For estimating heterogeneous SoS distribution, local SoS maps were reconstructed in~\cite{jaeger_2015} using steered plane waves and frequency domain reconstruction. 
A spatial domain reconstruction approach in \cite{sanabria_2018} allowed successful image quality that can be used in multiple clinical applications, such as for breast cancer diagnosis~\cite{Ruby_breast_19}.
In~\cite{rau_2021}, instead of plane waves, diverging wave transmission is proposed to minimize wavefront diffraction artifacts in SoS reconstruction. 
These above local SoS mapping methods all utilize beamformed images and thus require an initial SoS assumption; hence their results also highly depend on the accuracy of such assumed SoS value.

\section{Method}
\begin{figure}[b]
\centering
\includegraphics[width=0.94\linewidth]{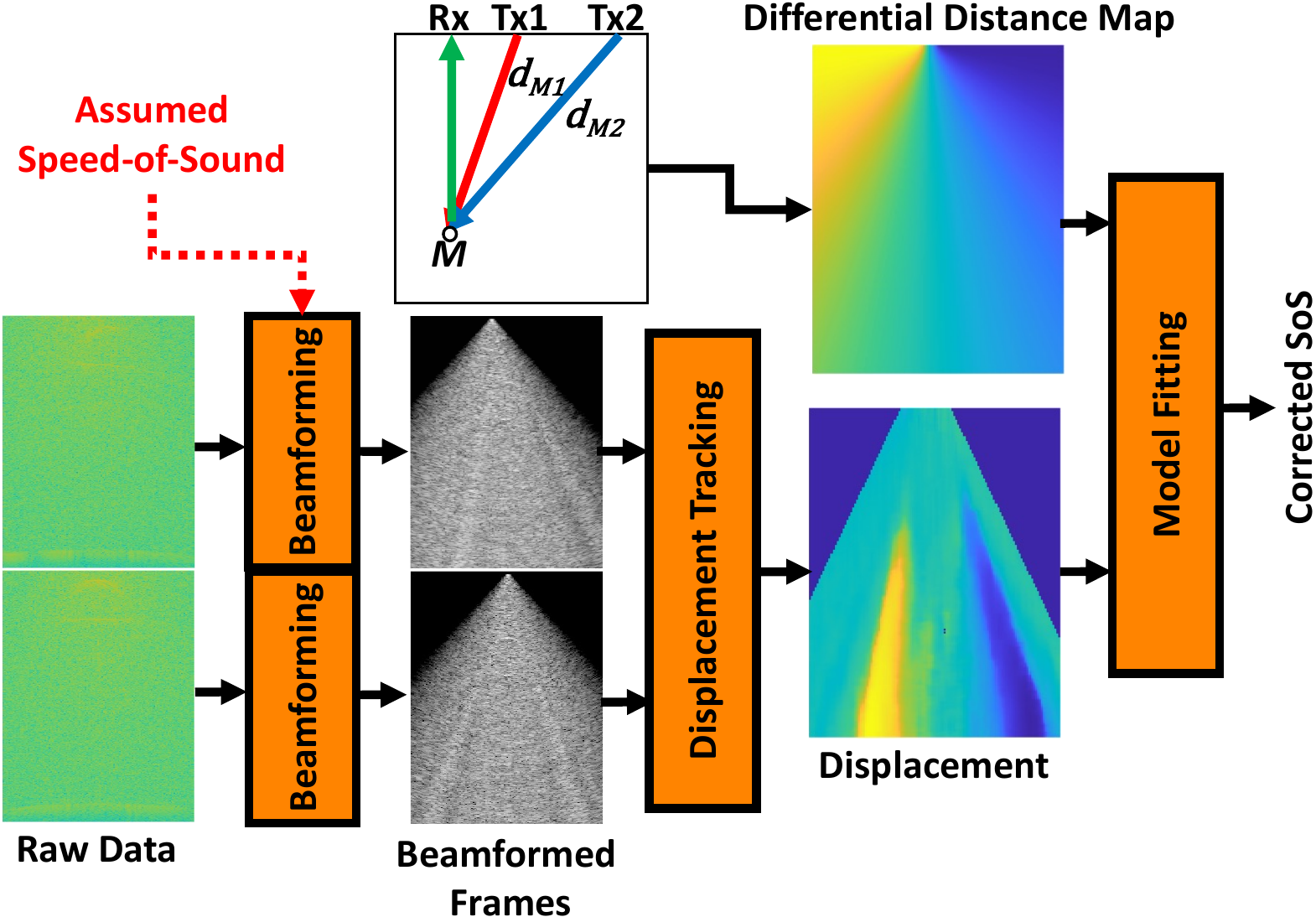}
\caption{Processing overview of the proposed method.}
\label{fig:processing_pipeline}
\end{figure}

In this work, we devise an analytical method for mean SoS estimation using the geometric disparities between images generated with different transmission/receive events during data acquisition.
An overview of our processing pipeline is shown in Fig.\,\ref{fig:processing_pipeline}.
In our processing, we first beamform the received temporal raw data using an arbitrary assumed SoS value. 
Any disparity between the assumed and actual SoS then causes a spatially-varying shift in the beamformed frames. 
We estimate these shifts between the beamformed RF frames using a motion-estimation (displacement-tracking) method. 
We devise an algebraic model relating such speckle shifts to differential Tx distances as well as the disparity between the actual and beamforming SoS values.
Using this model, we then calculate a correction factor to update the assumed value with an accurate SoS estimate.

\subsection{Analytical model}
\begin{figure}
\centering
\includegraphics[width=0.48\linewidth]{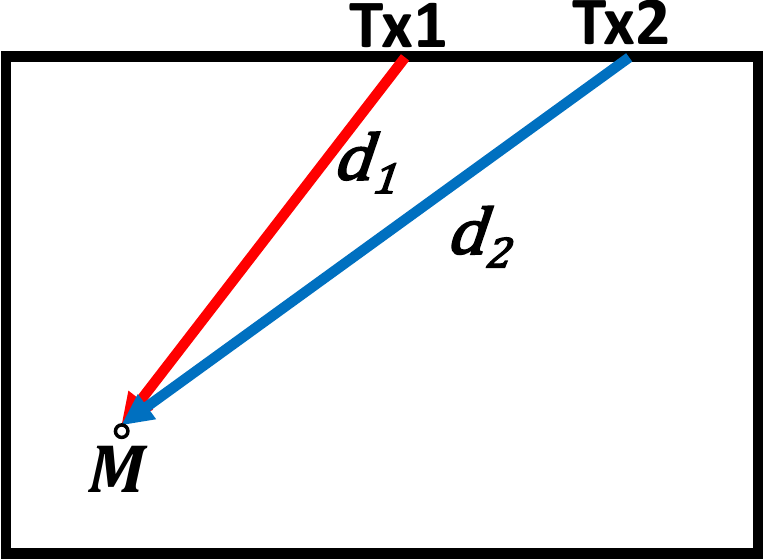}
\caption{Illustration of the transmission geometry.}
\label{fig:BF_grid_illustration}
\end{figure}

Consider two transmits Tx1 and Tx2 with different distances to a point $M$, as shown in Fig.\,\ref{fig:BF_grid_illustration}. 
By using dynamic receive aperture (with an F-number of N), we place the Rx apertures directly above the beamformed point, such that only the Tx paths differ for different transmits. 
Any disparity between the unknown actual SoS $c_\mathrm{ac}$ and assumed beamforming SoS $c_\mathrm{bf}$ is expected to cause a spatial shift between different beamformed frames.
Using transmission geometry, this shift can be written as a spatial function of $\left(\frac{c_{\mathrm{bf}}}{c_{\mathrm{ac}}}-\frac{c_{\mathrm{ac}}}{c_{\mathrm{bf}}}\right)$ as follows:
\begin{equation} 
\label{eq:shift_differential}
\Delta x =  x_2 - x_1 =  \left(\frac{d_2-d_1}{2} \right)\left(\frac{c_\mathrm{{bf}}}{c_\mathrm{{ac}}}-\frac{c_\mathrm{{ac}}}{c_\mathrm{{bf}}}\right).
\end{equation}
where $\Delta x$ is the expected shift between image features observed at locations $x_1$ and $x_2$ in the two beamformed frames from different Tx events.
Herein $d_1$ and $d_2$ are known physical distances of the beamformed image point to each Tx element.
Note that their difference ($d_1$$-$$d_2$) changes spatially, and in particular laterally, across the image frame.
Aggregating \eqref{eq:shift_differential} for every beamformed point with corresponding geometric distances, we  arrive at the following linear system of equations:
\begin{equation} 
\label{eq:SoS_correction_linear}
\Delta{\mathbf{x}} = \mathbf{D} \gamma\ , \qquad \mathrm{where} \quad \gamma = \left(\frac{c_{\mathrm{bf}}}{c_{\mathrm{ac}}}-\frac{c_{\mathrm{ac}}}{c_{\mathrm{bf}}}\right)
\end{equation}
with $\gamma$ the correction factor, $\Delta{\mathbf{x}}$ the displacement map, and $\mathbf{D}$ the differential distance map, as illustrated in Fig.\,\ref{fig:processing_pipeline} (top-right). 
We use \eqref{eq:SoS_correction_linear} as an analytical model to predict global SoS.

\subsection{Model fitting to predict the global SoS}
Given imaging data, we measure the spatial distribution of point-wise shifts between beamformed frames using a speckle tracking algorithm, herein a time-delay estimation method based on normalized cross-correlation~\cite{reza_disp}. Finding the correction factor $\gamma$ can then be defined as an optimization problem (\emph{model fitting}) by minimizing the following cost function:
\begin{equation} 
\label{eq:SoS_correction_problem}
\hat{\gamma} = \arg\min_\gamma || \Delta{\mathbf{x}} - \gamma \mathbf{D} ||_2\,.
\end{equation}
SoS can then be updated to a corrected value using:
\begin{equation} 
\label{eq:corrected_SoS_calculation}
c_\mathrm{{corrected}} = \frac{c_\mathrm{{bf}} (\sqrt{\hat{\gamma}^2 +4}-\hat{\gamma})}{2}
\end{equation}

\subsection{Application on local SoS reconstruction}
An incorrect SoS assumption does not only deteriorate B-mode imaging, but also negatively affects local SoS reconstruction process due to the use of beamformed images therein.
To demonstrate this and the use of our proposed method to this end, we first briefly explain local SoS reconstruction procedure below.
For tomographic reconstruction of SoS maps, first a differential path matrix $\mathbf{L}$ with each row indicating the integral distances from Tx and Rx elements to a spatial image location is formed as in~\cite{rau_2021}. 
Then, given differential displacement measurements ${\Delta\mathbf{\tau}}$, reconstructing a spatial slowness map $\hat{\sigma}$ (inverse of SoS) is formulated as:
\begin{equation}
    \label{eq:SoS_Image_Reconstruction}
    \hat{\sigma} = \argmin_\sigma || \mathbf{L} (\sigma - \sigma_0) -{\Delta\mathbf{\tau}}||_1 + \lambda ||\mathbf{B}\sigma||_1,
\end{equation}
where $\sigma_0$ is the initial slowness, $\mathbf{B}$ is the regularization matrix, and $\lambda$ is the trade-off parameter between data and regularization terms. Following~\cite{rau_2021}, we use total variation regularization with anisotropically weighted spatial gradients and solve the inverse problem using the limited memory Broyden-Fletcher-Goldfarb-Shanno (L-BFGS) algorithm.

\section{Results}

We conducted experiments with numerical phantoms simulated using \textit{k-Wave} toolbox~\cite{k_Wave}. 
We simulated a linear transducer with $128$ channels and $300$\,mm pitch, on a 2D phantom of size 40x55\,mm. 
Spatial and temporal resolutions were set to be $75\,\mu$m and $6.25$\,ns, respectively. 
Utilizing diverging waves~\cite{rau_2021}, for each Tx only a single is activated.
For each Tx pulse, we used 4 half cycles with a center frequency of $f_c$$=$$5$\,MHz. 
We study errors due to over- and under-assumption of beamforming SoS values, by initially beamforming the raw channel Rx data with 8 different global SoS values of \{1,\,3,\,5,\,10\}\% higher and lower than the known ground-truth SoS.
We then employed our proposed method to evaluate the corrected SoS values.
Without loss of generality, in this paper for SoS correction we used the beamformed RF frame pair from Tx of channels 55 and 65 only. 

\subsection{Global SoS estimation in homogeneous medium}

\begin{figure}%
\centering%
\begin{tabular}{@{}c@{\quad}c@{}}%
\includegraphics[height=0.47\linewidth]{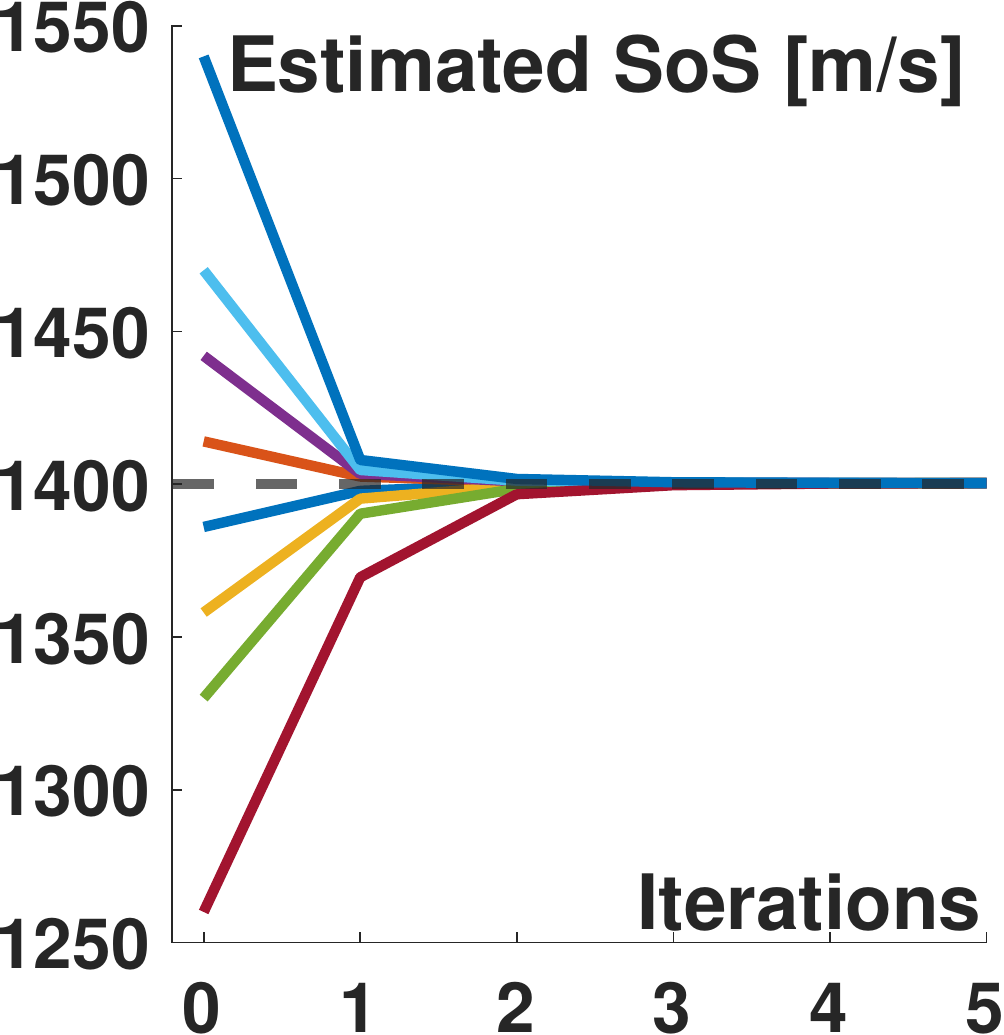}   &              \includegraphics[height=0.47\linewidth]{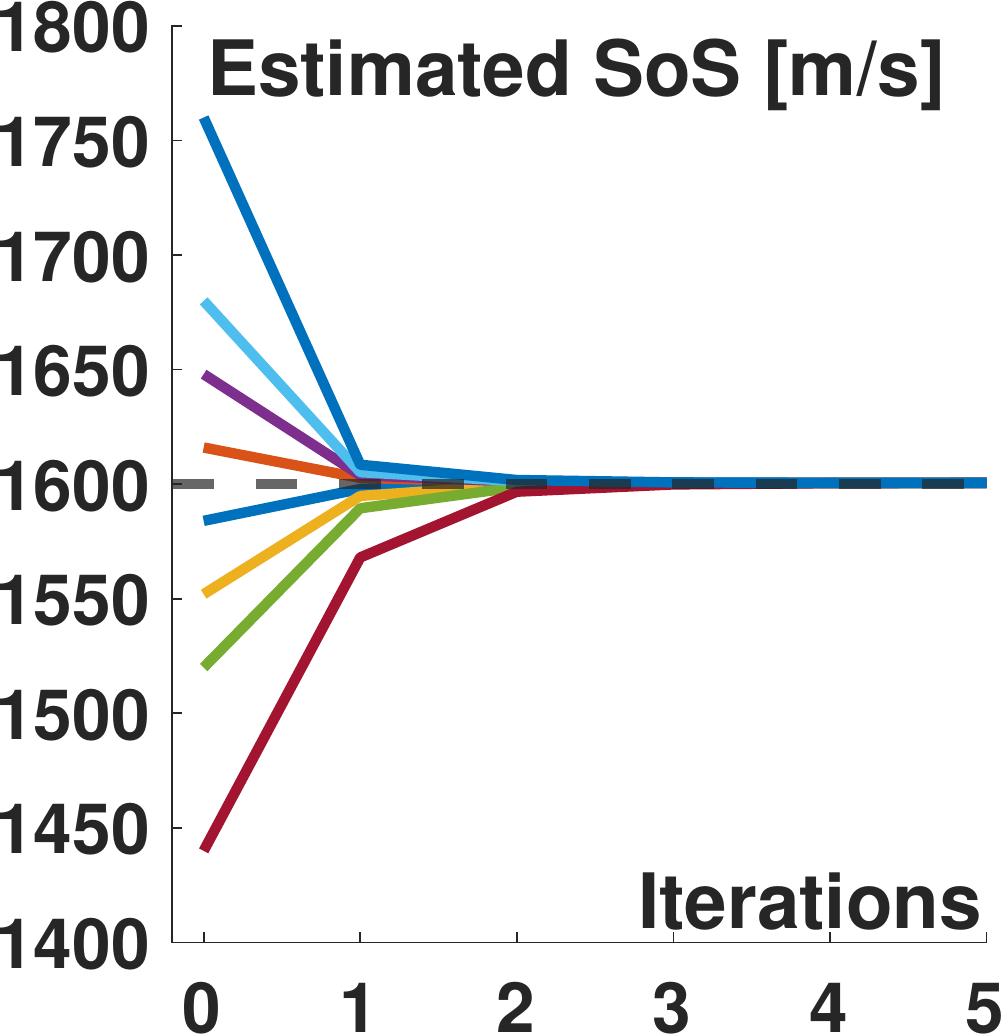}\\ 
        (a) 1400 [m/s] & (b) 1600 [m/s]
    \end{tabular}
        \caption{Iterative estimation of global SoS values from different initial SoS assumptions for homogeneous phantoms with SoS values of 1400 and 1600\,m/s.}
        \label{fig:convergence_homogenous}
    \end{figure}

First we studied accuracy in two homogeneous numerical phantoms with SoS values of 1400 and 1600\,m/s, where internal scattering was simulated using minor density variations. 
Given 8 different initial SoS assumptions, Fig.\,\ref{fig:convergence_homogenous} shows the SoS values corrected by our method, indicating a convergent pattern when applied iteratively.
The ground-truth SoS values are found within few iterations, regardless of the initial estimate. 

\subsection{Global SoS estimation in heterogeneous medium}
Next we studied the feasibility of our method in heterogeneous phantoms.
To that end, we utilized the set of 32 numerical phantom simulations described in~\cite{melanie_2020}.
For a heterogeneous phantom, since there is no ground-truth global SoS value, we conduct this evaluation via comparisons against the \emph{mean} SoS value within each phantom, which vary across 32 phantoms depending on their inclusion SoS and size.
Accordingly, we also set the initial assumed beamforming SoS to \{1,\,3,\,5,\,10\}\% higher and lower than the phantom mean.
For each phantom, we then computed mean absolute error (MAE) between the mean of ground-truth phantom SoS map and our global SoS prediction (or the assumed value for the initial step).
Table~\ref{tab:improvement_in_mean_SoS} presents average MAEs over 32 inclusion phantoms, for different initial beamforming SoS assumptions and their corrected values after the 1st and the 5th correction iterations using our method. 
\begin{table*}
\renewcommand{\arraystretch}{1.3}
\caption{Mean Absolute Error (MAE) values [m/s] averaged over 32 heterogeneous numerical phantoms.}
\label{tab:improvement_in_mean_SoS}
\centering
\scalebox{1}{
\begin{tabular}{l|cccc|cccc}
\hline
\multicolumn{1}{l}{\textbf{Experimental Setting}} & \multicolumn{4}{c}{\textbf{Under-Assumption}}&\multicolumn{4}{c}{\textbf{Over-Assumption}}\\
\cmidrule(lr){2-5}\cmidrule(lr){6-9}
\multicolumn{1}{l}{\textbf{ }} & \textbf{1\%}&\textbf{3\%}&\textbf{5\%}&\textbf{10\%}&\textbf{1\%}&\textbf{3\%}&\textbf{5\%}&\textbf{10\%}\\
\hline
Error for assumed beamforming SoS
& 15.02&45.07&75.12&150.23 
&15.02&45.07&75.12&150.23 \\
Corrected using geometric model from~\cite{xenia_2021}
&9.47&37.20&57.66&92.68 
&11.37&34.17&54.66&73.27\\
Corrected using our proposed model (iteration\,1)
&5.25&6.33&13.10&43.42 
&3.90&4.25&5.97&26.25 \\
\textbf{Corrected using our proposed model (iteration\,5)} 
&4.54&4.55&4.57&4.57 
&4.52&4.52&4.52&4.49\\
\hline
\textbf{Improvement [\%] from assumed to our corrected}&69.8&89.9&93.9&97.0
&69.9&90.0&94.0&97.0\\
\hline
\end{tabular}}
\end{table*}
We also compared our method favorably to global SoS values corrected using the geometric model from~\cite{xenia_2021}, which is an inexact model which was remedied in that work using a tedious and potentially error-prone calibration step.
As seen in Table~\ref{tab:improvement_in_mean_SoS}, our proposed method successfully reduces the average MAE under 5\,m/s after five iterations, both from over- and under-assumption; leading to MAE improvements over 30 folds in some experimental settings.
The remaining MAE can be explained by the fact that mean phantom SoS used for comparison here is indeed not a ground-truth, but is rather a surrogate, since the SoS location, Tx/Rx geometry, etc may all affect how an SoS heterogeneity could factor into beamforming.

\begin{figure}%
\centering%
\begin{tabular}{@{}cccc@{}}%
\includegraphics[height=0.28\linewidth]{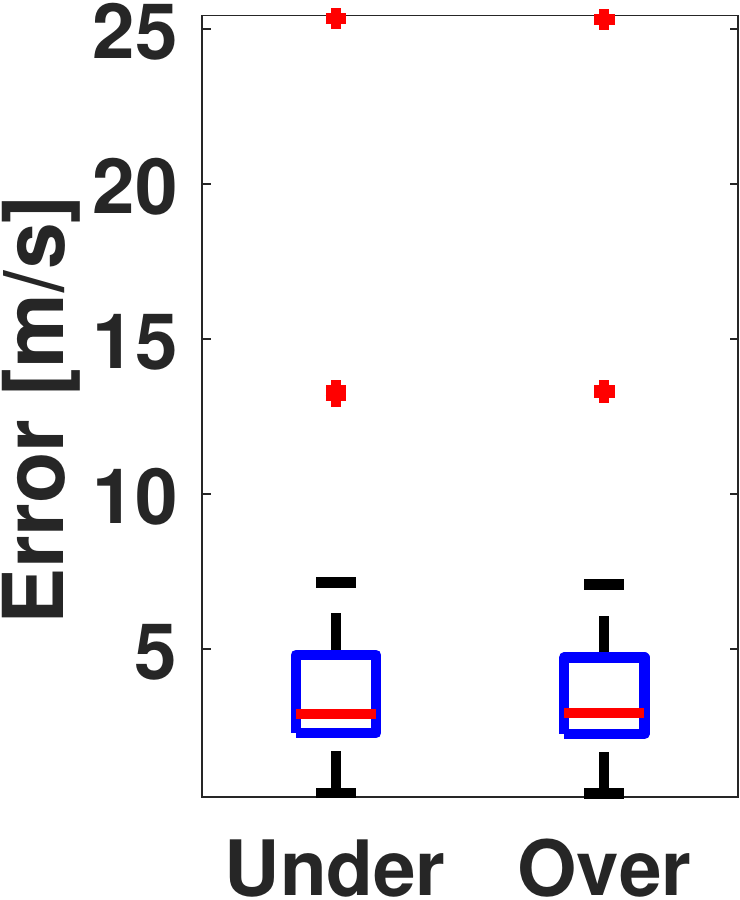}   &         \includegraphics[height=0.28\linewidth]{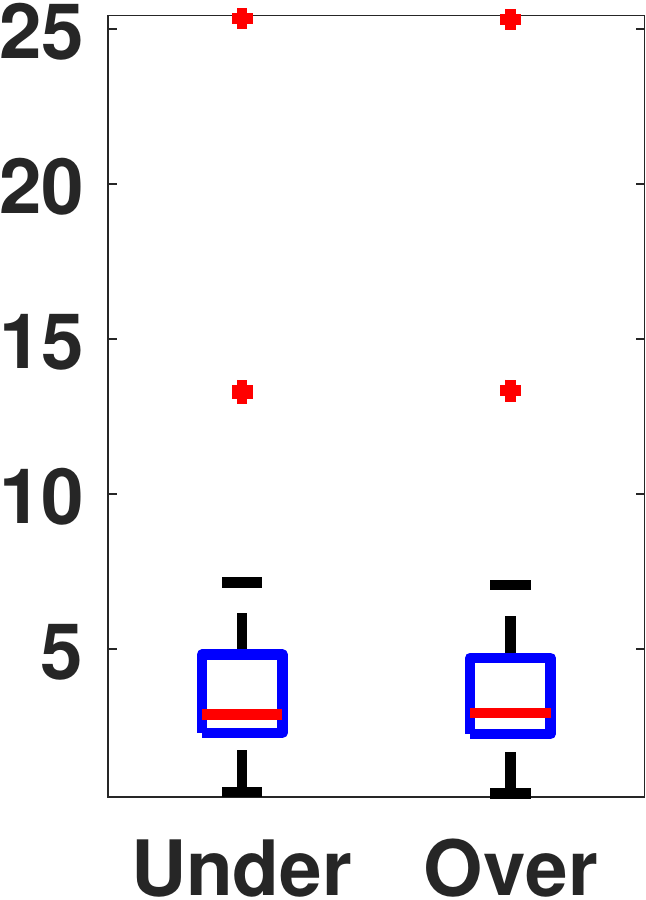} &
    \includegraphics[height=0.28\linewidth]{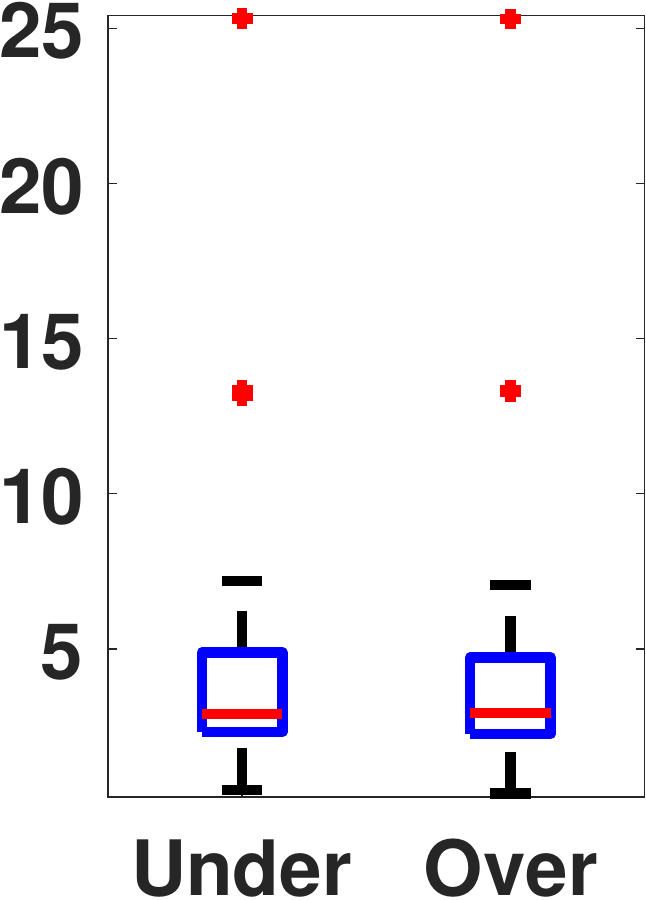} &
     \includegraphics[height=0.28\linewidth]{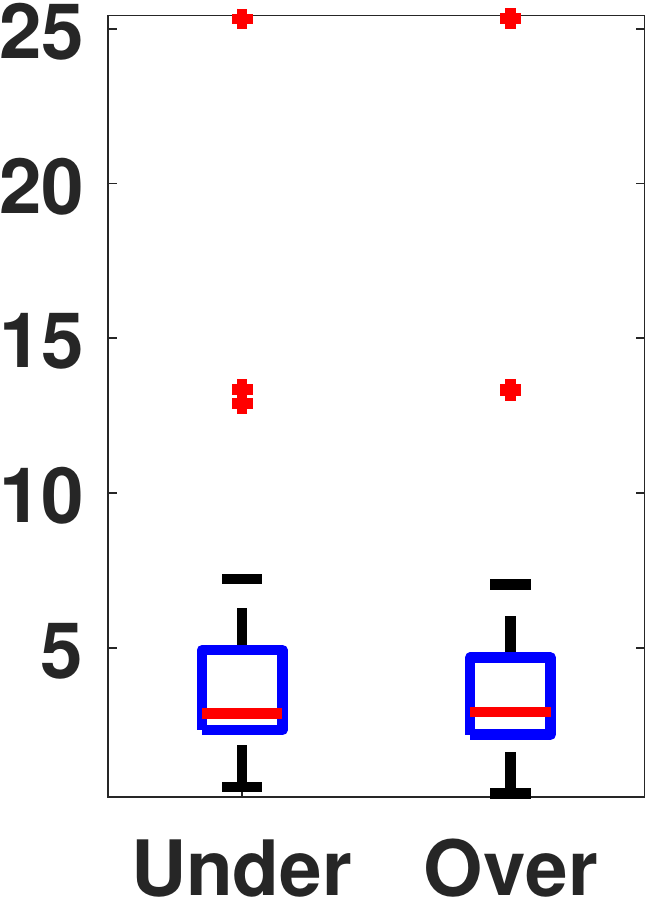}\\ 
        (a) 1\% & (b) 3\% & (c) 5\%  & (d) 10\%
    \end{tabular}
        \caption{Distributions of MAEs between corrected SoS (iteration\,5) from mean ground-truth SoS for each phantom for different initial under- and over-assumption percentages.}
        \label{fig:MAE_box}
    \end{figure}

Distributions of differences of the corrected SoS from the mean phantom SoS are presented in Fig.\,\ref{fig:MAE_box} for different experimental settings.
For each phantom, the corrected SoS is seen to converge similar values regardless of the initialization, which indicates the robustness of our analytical approach.

\subsection{Effect of global SoS assumption in local SoS reconstruction}
\begin{figure}
      \centering
        \begin{subfigure}[b]{0.49\linewidth}
            \centering
           \includegraphics[width=\linewidth]{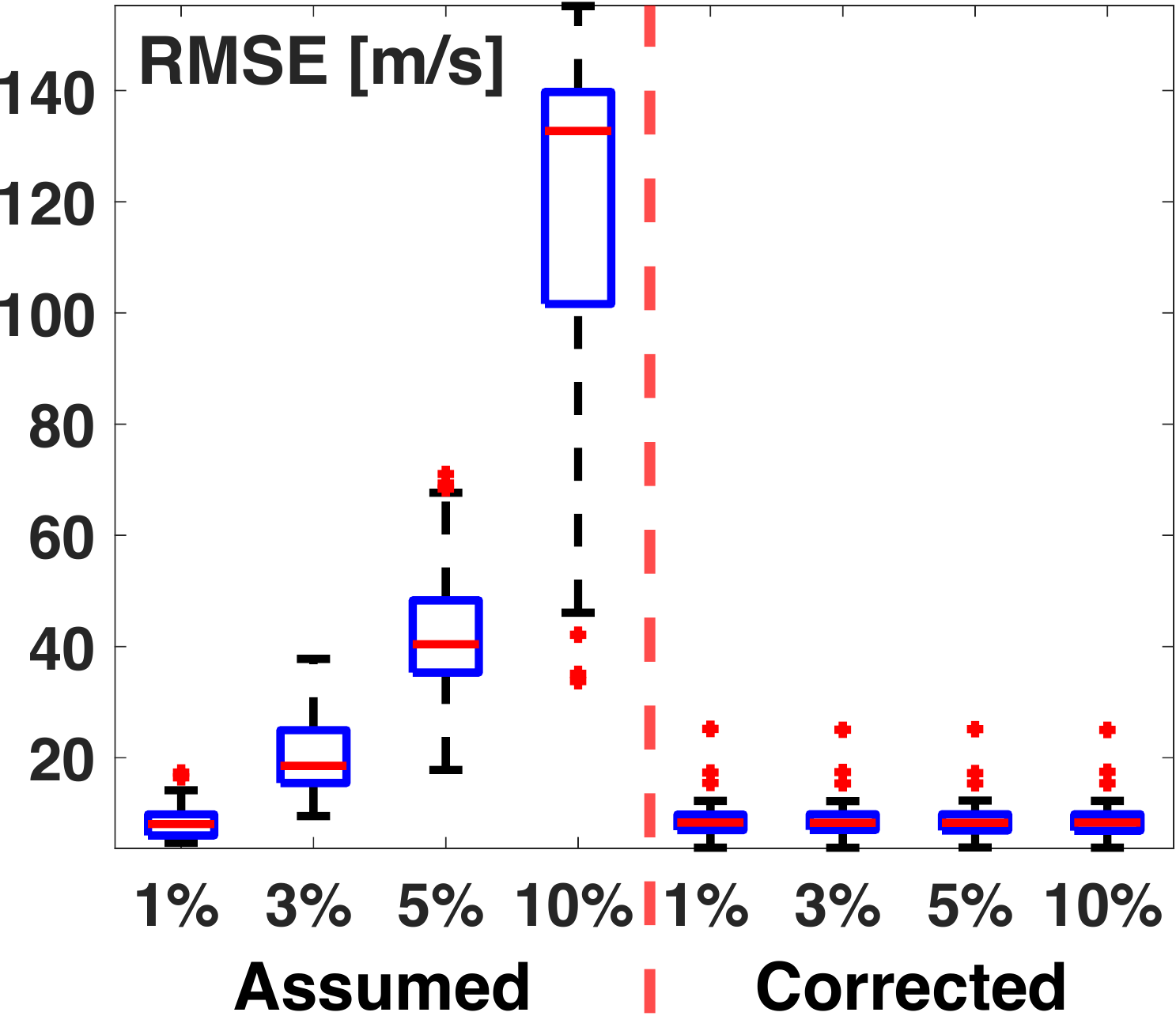}

            \caption{Under-assumption}
        \end{subfigure}
        \hfill
        \begin{subfigure}[b]{0.49\linewidth}  
            \centering 
            \includegraphics[width=\linewidth]{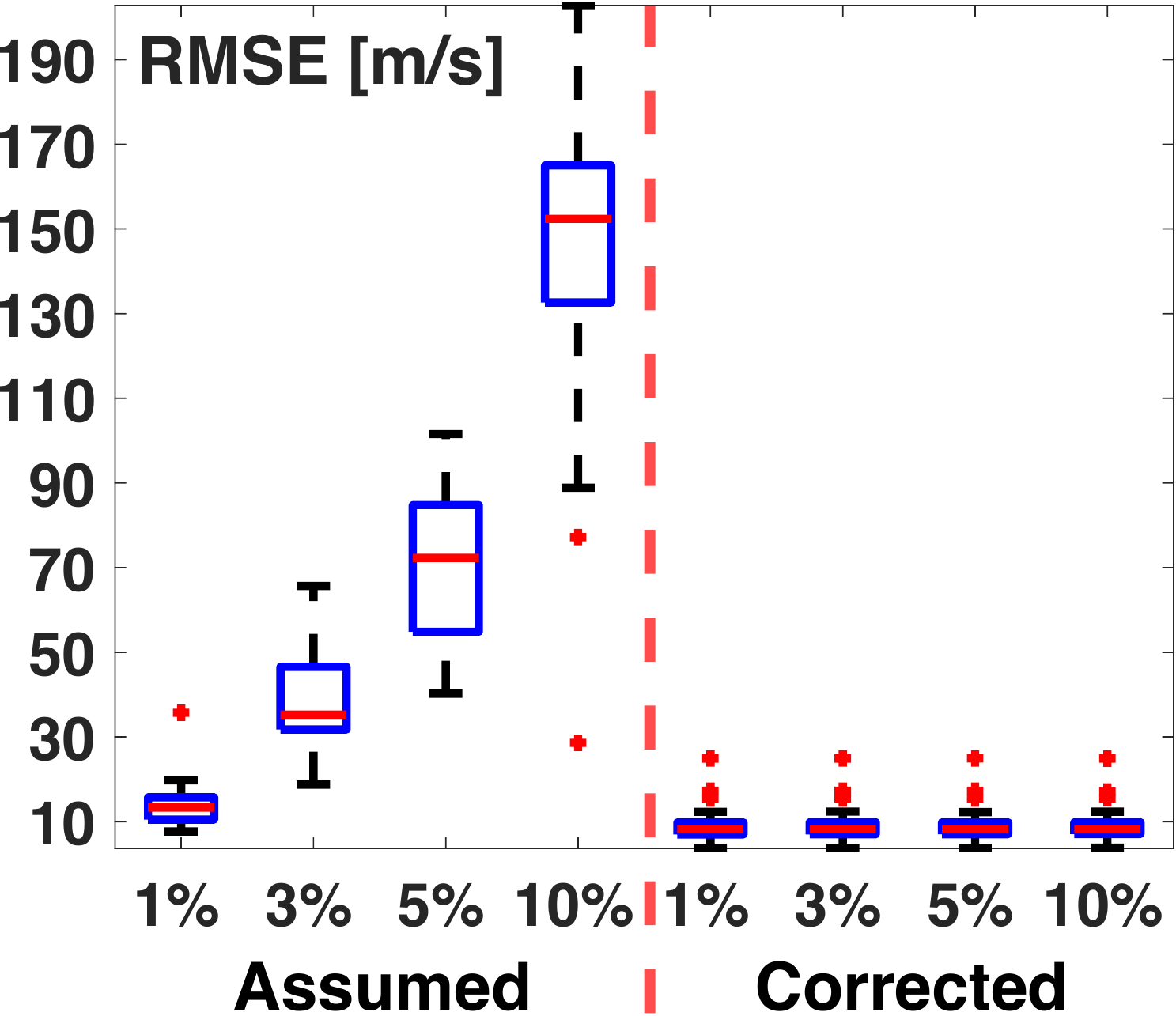}
            \caption{Over-assumption}
        \end{subfigure}
        \caption{Distributions of RMSEs for 32 numerical phantoms for different beamforming SoS (a)~under- and (b)~over-assumption percentages and their corrections.}
        \label{fig:RMSE_box}
    \end{figure}

Since local SoS reconstruction uses displacements between beamformed RF frames, its accuracy also relies on a good beamforming SoS assumption.
To study this, we compared local SoS reconstructions of the 32 phantoms, using under-/over-assumed beamforming SoS values and SoS values corrected by different methods. 
For local SoS reconstructions, we used 6 pairs of beamformed RF frames as in~\cite{melanie_2020}.
For each reconstruction, we computed root mean square error (RMSE) with respect to the known ground truth SoS map.
Average RMSEs over 32 numerical phantoms are given in Table~\ref{tab:RMSE_values} for each method for beamforming SoS determination. 
\begin{table*}[t]
\renewcommand{\arraystretch}{1.3}
\caption{Root Mean Square Error (RMSE) values [m/s] averaged over 32 heterogeneous numerical phantoms.}
\label{tab:RMSE_values}
\centering
\scalebox{1}{
\begin{tabular}{l|cccc|cccc}
\hline
\multicolumn{1}{l}{\textbf{Experimental Setting}} & \multicolumn{4}{c}{\textbf{Under-Assumption}}&\multicolumn{4}{c}{\textbf{Over-Assumption}}\\
\cmidrule(lr){2-5}\cmidrule(lr){6-9}
\multicolumn{1}{l}{\textbf{ }} & \textbf{1\%}&\textbf{3\%}&\textbf{5\%}&\textbf{10\%}&\textbf{1\%}&\textbf{3\%}&\textbf{5\%}&\textbf{10\%}\\
\hline
Error for assumed beamforming SoS
&8.54&20.23&42.37&115.75
&13.87&39.08&70.16&143.82\\
Corrected using geometric model from~\cite{xenia_2021}
&13.12&34.37&52.23&87.81
&9.39&14.34&25.76&42.25\\
Corrected using our proposed model (iteration\,1)
&8.60&8.57&8.51&20.00 
&9.27&9.37&9.85&24.69 \\
\textbf{Corrected using our proposed model (iteration\,5)} 
&9.16&9.16&9.11&9.09 
&9.13&9.13&9.12&9.12\\
\hline
\textbf{Improvement [\%] from assumed to our corrected}&-7.3&54.7&78.5&92.2
&34.2&76.6&87.0&93.7\\
\hline

\end{tabular}}
\end{table*}
As seen, our method significantly improves the accuracy of tomographic SoS reconstructions, and more importantly make the results of this quantitative imaging method robust to and independent of the unknown global SoS value for beamforming. 
In particular, for 5\% initial offset, RMSE is improved by 78.5\% for under-assumption and 87.0\% for over-assumption cases. 
Distributions of RMSEs for each phantom reconstructed using initial assumed and 5th-iteration corrected SoS values are presented in Fig.\,\ref{fig:RMSE_box}, demonstrating a consistent improvement of SoS reconstructions for all phantoms regardless of the initial SoS assumption error. 

\begin{figure}
\centering
\includegraphics[width=\linewidth]{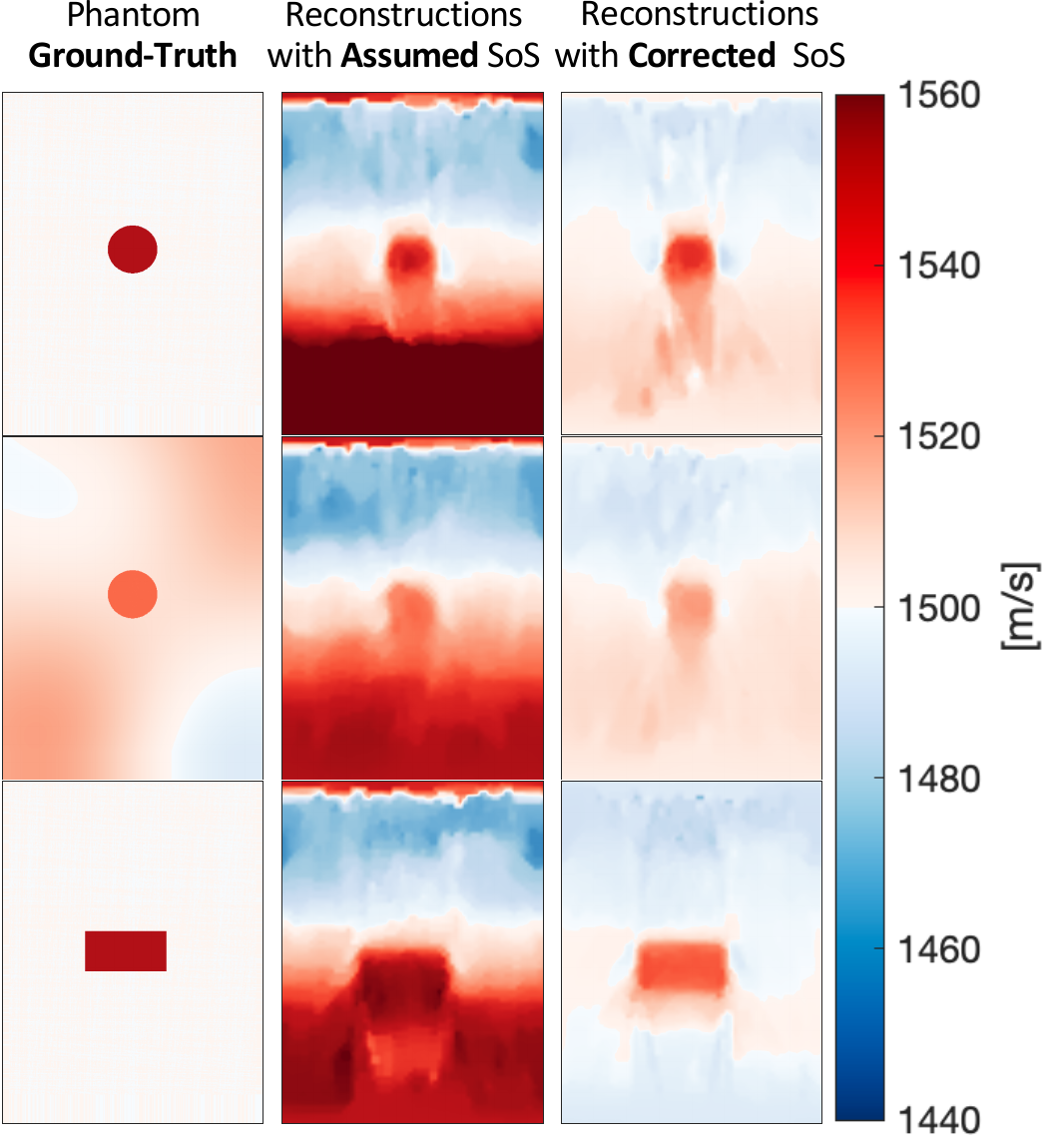}
\caption{Local SoS reconstructions for 3 sample numerical phantoms, with 3\% error in beamforming SoS assumption and with our corrected beamforming SoS values.}
\label{fig:SoS_Recons_3_percent}
\end{figure}

In Fig.\,\ref{fig:SoS_Recons_3_percent}, we illustrate three sample phantoms and their reconstructions for 3\% assumption error and corrected SoS values. 
As seen, after the correction,  reconstruction are also substantially improved qualitatively. 
Specifically, background SoS values and deeper regions are better reconstructed, where the inclusions can be distinguished more clearly.

\section{Conclusion}
In this work, we have proposed an analytical method for the estimation of beamforming SoS by utilizing the geometric disparities between different transmits. 
We have devised a model for the differential spatial shifts between different beamformed RF frames, and use this in a closed-form model fitting approach to estimate a corrected beamforming SoS from any initial arbitrary SoS assumption.
By applying this iteratively, we show that actual SoS values are estimated robustly and accurately.
We have studied our method on numerically simulated phantoms.
We have shown the accuracy of our method and its behaviour with iterative use on homogeneous phantoms, 
and also demonstrated its utility in significantly improving local SoS map reconstruction accuracy.
Our method utilizes an improved model compared to~\cite{xenia_2021}; relies only on the known transducer geometry and hence does not require any prior training or calibration; and can be used iteratively for increased accuracy.

\bibliographystyle{IEEEtran}
\bibliography{global_SoS}

\begin{thebibliography}{10}
\providecommand{\url}[1]{#1}
\csname url@samestyle\endcsname
\providecommand{\newblock}{\relax}
\providecommand{\bibinfo}[2]{#2}
\providecommand{\BIBentrySTDinterwordspacing}{\spaceskip=0pt\relax}
\providecommand{\BIBentryALTinterwordstretchfactor}{4}
\providecommand{\BIBentryALTinterwordspacing}{\spaceskip=\fontdimen2\font plus
\BIBentryALTinterwordstretchfactor\fontdimen3\font minus
  \fontdimen4\font\relax}
\providecommand{\BIBforeignlanguage}[2]{{%
\expandafter\ifx\csname l@#1\endcsname\relax
\typeout{** WARNING: IEEEtran.bst: No hyphenation pattern has been}%
\typeout{** loaded for the language `#1'. Using the pattern for}%
\typeout{** the default language instead.}%
\else
\language=\csname l@#1\endcsname
\fi
#2}}
\providecommand{\BIBdecl}{\relax}
\BIBdecl

\bibitem{rau_2021}
R.~Rau, D.~Schweizer, V.~Vishnevskiy, and O.~Goksel, ``Speed-of-sound imaging
  using diverging waves,'' \emph{International journal of computer assisted
  radiology and surgery}, vol.~16, pp. 1201--1011, 2021.

\bibitem{anderson_1998}
M.~E. Anderson and G.~E. Trahey, ``The direct estimation of sound speed using
  pulse-echo ultrasound,'' \emph{The Journal of the Acoustical Society of
  America}, vol. 104, pp. 3099--106, 1998.

\bibitem{byram_2012}
B.~Byram, G.~Trahey, and J.~Jensen, ``A method for direct localized sound speed
  estimates using registered virtual detectors,'' \emph{Ultrasonic imaging},
  vol.~34, pp. 159--180, 2012.

\bibitem{shen_2020}
C.-C. Shen and K.-L. Tu, ``Ultrasound {DMAS} beamforming for estimation of
  tissue speed of sound in multi-angle plane-wave imaging,'' \emph{Applied
  Sciences}, vol.~10, no.~18, p. 6298, 2020.

\bibitem{xenia_2021}
X.~Augustin, L.~Zhang, and O.~Goksel, ``Estimating mean speed-of-sound from
  sequence-dependent geometric disparities,'' in \emph{2021 IEEE International
  Ultrasonics Symposium (IUS)}, 2021, pp. 1--4.

\bibitem{jaeger_2015}
M.~Jaeger, G.~Held, S.~Preisser, S.~Peeters, M.~Grünig, and M.~Frenz,
  ``Computed ultrasound tomography in echo mode for imaging speed of sound
  using pulse-echo sonography: proof of principle,'' \emph{Ultrasound Med
  Biol.}, vol.~41, pp. 235--250, 2015.

\bibitem{sanabria_2018}
S.~J. Sanabria, E.~Ozkan, M.~B. Rominger, and O.~Goksel, ``Spatial domain
  reconstruction for imaging speed-of-sound with pulse-echo ultrasound:
  simulation and in vivo study,'' \emph{Physics in medicine and biology},
  vol.~63, p. 215015, 2018.

\bibitem{Ruby_breast_19}
L.~Ruby, S.~J. Sanabria, K.~Martini, K.~J. Dedes, D.~Vorburger, E.~Oezkan,
  T.~Frauenfelder, O.~Goksel, and M.~B. Rominger, ``Breast cancer assessment
  with pulse-echo speed of sound ultrasound from intrinsic tissue reflections:
  Proof-of-concept,'' \emph{Investigative Radiology}, vol.~54, no.~7, pp.
  419--427, Jul 2019.

\bibitem{reza_disp}
R.~Z. Azar, O.~Goksel, and S.~E. Salcudean, ``Sub-sample displacement
  estimation from digitized ultrasound rf signals using multi-dimensional
  polynomial fitting of the cross-correlation function,'' \emph{IEEE Trans
  Ultras, Ferroelec, Freq Control}, vol.~57, no.~11, pp. 2403--2420, 2010.

\bibitem{k_Wave}
B.~E. Treeby and B.~T. Cox, ``{k-Wave: MATLAB toolbox for the simulation and
  reconstruction of photoacoustic wave fields},'' \emph{Journal of Biomedical
  Optics}, vol.~15, no.~2, pp. 1 -- 12, 2010.

\bibitem{melanie_2020}
M.~Bernhardt, V.~Vishnevskiy, R.~Rau, and O.~Goksel, ``Training variational
  networks with multidomain simulations: Speed-of-sound image reconstruction,''
  \emph{IEEE Transactions on Ultrasonics, Ferroelectrics, and Frequency
  Control}, vol.~67, no.~12, pp. 2584--2594, 2020.

\end{thebibliography}

\end{document}